\documentclass[aps,twocolumn,amsmath,floatfix,byrevtex,prb,nobibnotes]{revtex4}
\usepackage{graphicx}
\usepackage{amssymb}
\usepackage{bm}

\begin{document}
\title{Interference effects in the counting statistics of electron transfers through a double quantum dot}

\author{Sven Welack}
\thanks{also at: Department of Chemistry, Hong Kong University of Science and Technology, Kowloon, Hong Kong (SAR)}
\email{swelack@uci.edu}
\author{Massimiliano Esposito}
\thanks{also at: Center for Nonlinear Phenomena and Complex Systems, Universite Libre de Bruxelles, Code Postal 231, Campus Plaine, B-1050 Brussels, Belgium.}
\author{Upendra Harbola}
\author{Shaul Mukamel}\email{smukamel@uci.edu}
\affiliation{Department of Chemistry, University of California, Irvine, California 92697, USA.}

\today
\begin{abstract}
We investigate the effect of quantum interferences and Coulomb interaction on the counting statistics of electrons crossing a double quantum dot in a parallel geometry using a generating function technique based on a quantum master equation approach. The skewness and the average residence time of electrons in the dots are shown to be the quantities most sensitive to interferences and Coulomb coupling. The joint probabilities of consecutive electron transfer processes show characteristic temporal oscillations due to interference. The steady-state fluctuation theorem which predicts a universal connection between the number of forward and backward transfer events is shown to hold even in the presence of Coulomb coupling and interference.
\end{abstract}

\maketitle

\section{Introduction}

Fast and sensitive charge detectors and highly-stable current bias sources have made it possible to measure individual electrons crossing arrays of tunnel junctions\cite{Byla04} or quantum dots\cite{Lu03, Fuji04, Gustav06} (QDs). Directional forward and reverse counting through two quantum dots in a series has been reported\cite{Fuji06}. Spurred by this experimental progress, electron counting statistics (ECS) in nanosystems has attracted recent theoretical interest both in the noninteracting\cite{Levi04,Blant00,Gurv97,Wabnig05, Rammer04, Shela03, Flindt05, Kiess06,Utsumi06,groth07,max06a} and Coulomb blockade regime\cite{wang07,Bagr03}. It has been shown that strong Coulomb interactions suppress large current fluctuations\cite{Bagr03}. 

Double quantum dot (DQD) systems in parallel geometry\cite{chen04,lopez02,goldstein07,tanaka05} and single multi-level quantum dots\cite{buesser04,max06} can display interference effects due to the multiple paths that electrons take to cross the junction when Coulomb interaction is present. These effects on the average electric current and population \cite{chen04,lopez02,goldstein07,tanaka05,buesser04,max06} do not require a magnetic field as in double quantum dot Aharonov-Bohm interferometers\cite{utsumi04,bai04,kubala02,tokura07,boese02,kang04}.
A recent study of interference effects on the electron transfer statistics has been shown that they can induce super-Poisson shot noise in Coulomb blockade regime and the high bias limit\cite{wang07}. Using the terminology of\cite{wang07}, we denote the couplings between populations and coherences in the many-body eigenbasis as interferences. 

In this paper we extend full counting statistics to arbitrary Coulomb coupling in a DQD junction by employing a quantum master equation approach (QME). We find that the third moment and the average electron residence time in the system provide useful indicators for interferences effects on the ECS. We also introduce joint elementary probabilities to electron transport which can be obtained from a time-series analysis of single transfer events. These probabilities reveal temporal oscillations induced by interferences. Their amplitude is amplified by the Coulomb coupling and the frequency is determined by the energy detuning between the orbitals of the dots. This method can reveal detunings much smaller than $k_b T$. 
The employed elementary probabilities are equivalent to waiting time distributions recently investigated in single electron transport\cite{brandes07} and electron transport through single molecules\cite{sven07_3}.

ECS experiments have been proposed recently\cite{max06a} as good candidates to test the validity in the quantum domain of far-from-equilibrium fluctuation relations, called fluctuation theorems (FTs), which have raised attention in classical systems\cite{Galla95,Jarz97,Lebo99,Searl99,seif05}. In Ref.\cite{max06a} noninteracting electrons were considered and interferences were absent. We demonstrate in this paper that the FT still holds in the presence of interferences and Coulomb repulsion.

Counting statistics in systems with interferences can be realized experimentally by connecting a DQD to two leads. The transitions between states can be measured by employing a quantum point contact\cite{Lu03, Fuji04, Gustav06}. As a reference point for the DQD junction, we also calculate the corresponding values for a QD with a single orbital where interferences are not possible. 

The paper is organized as follows: We present the DQD and QD models in sections (\ref{sec2}) and (\ref{sec1}), respectively. Equations of motion (EOM) for both models are derived in section (\ref{sec-master}). We then introduce the generating function (GF) in section (\ref{sec3}). In section (\ref{sec4}) we derive elementary probabilities for different electron detector configurations. The numerical results are presented in (\ref{sec5}) and we summarize and conclude in section (\ref{sec6}).

\subsection{Model A: Double quantum dot}\label{sec2}

\begin{figure}
\includegraphics[width=6cm,clip]{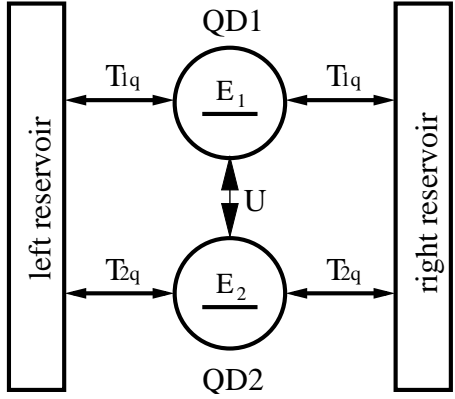}
\caption{Model A: Double quantum dot in junction, each quantum dot has a single spin orbital with energy $E_s$, connected to a left and right lead. The left and right coupling elements are equal and also $T_{1q}=T_{2q}$, only the Fermi energies of the left and right lead are different, thus defining a bias voltage $V=E_{F,l}-E_{F,r}$. $U$ is the Coulomb interaction parameter between the two dots.
\label{fig:2}}
\end{figure}

Model A consists of a DQD connected in parallel to two leads. Each QD contains a single spin orbital which is connected to the leads as shown in Fig.\,\ref{fig:2}. Experimentally it would correspond to spin polarized leads in an infinite magnetic field where only one spin state is accessible by electronic excitations. This spinless DQD model has been used to study the effect of interferences on average currents without\cite{lopez02,goldstein07,tanaka05} and with\cite{bai04,kang04,kubala02,tokura07} an additional Aharonov-Bohm phase.

The Hamiltonian
\begin{equation} \label{equ:Ham_total}
H=H_S+H_R+H_{SR},
\end{equation}
represents the system $H_S$, the lead $H_R$, and their coupling term $H_{SR}$.  We label the two dots with index $s=1,2$. In second quantization, the local-basis Hamiltonian of the system reads
\begin{equation}\label{equ:Ham_wire}
H_S = \sum_{s=1}^2 E_s \Psi_{s}^\dagger \Psi_{s} +  U \Psi_{1}^\dagger \Psi_{1} \Psi_{2}^\dagger \Psi_{2}.
\end{equation}
$U$ is the Coulomb coupling strength. The environment consists of two independent leads in thermal equilibrium. For each lead, the Hamiltonian is given by
\begin{equation} \label{equ:Ham_lead}
H_R=\sum_{q} \omega_q \Psi_{q}^\dagger \Psi_{q}
\end{equation}
with $\Psi_{q }^\dagger$ and $\Psi_{q}$ create and annihilate an electron in lead mode $\vert q  \rangle$ with energy $\omega_q$. To keep the notation simple, we will only refer to the left lead in further derivations. The right lead will be added to the final expressions. Since the leads are in thermal equilibrium, their occupation numbers are determined by Fermi-Dirac statistics
\begin{equation} \label{equ:equi2}
\langle \Psi_{q}^\dagger \Psi_{q'} \rangle_R = n_F(\omega_q-E_F) \delta_{qq'}
\end{equation}
where $n_F(\omega)=1/(e^{\beta \omega}+1)$ is the Fermi function, $\beta=1/kT$, and $E_F$ the Fermi energy. We denote the trace over the lead degrees of freedom by $\langle \cdot \rangle_R=\mathrm{tr}_R\lbrace \cdot \rho_R \rbrace$ where $\rho_R$ is the density operator of the lead. The coupling of the lead with the system can be written as
\begin{equation} \label{equ:Ham_coup}
H_{SR}= \sum_{sq} (T_{sq} \Psi_{s}^\dagger \Psi_q +T_{sq}^* \Psi_q^\dagger \Psi_{s}).
\end{equation}
$T_{sq}$ is the coupling strength between lead mode $q$ and the $s$-th QD. We assume weak system-lead coupling and no direct overlap (tunneling) between the wave functions of the left and right lead.

\begin{figure}
\includegraphics[width=6cm,clip]{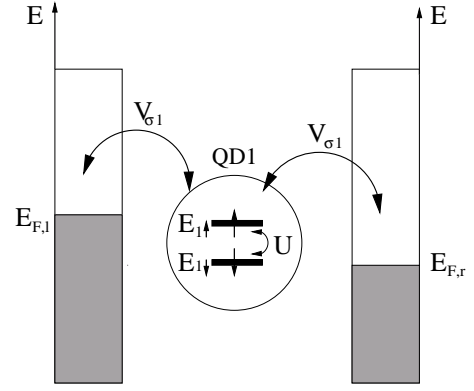}
\caption{Model B: A single orbital QD with energies $E_{1\uparrow}$, $E_{2\downarrow}$ for the spin-up and spin-down state. The coupled leads have Fermi energies $E_{F,l}$ and $E_{F,r}$. $U$ is the Coulomb coupling parameter. The left and right coupling elements are equal.
\label{fig:1}}
\end{figure}

\subsection{Model B: Single quantum dot}\label{sec1}

The second model, shown in Fig.\,\ref{fig:1}, consists of a QD with a single orbital which can accommodate two electrons with opposite spins coupled to two leads. In analogy to Model A, the total Hamiltonian can be written as $H=H_S+H_{R}+H_{SR}$. The system part is given by
\begin{equation}\label{equ:Ham_wire2}
H_S= \sum_{\sigma} E_{1\sigma} \Psi_{1\sigma}^\dagger \Psi_{1\sigma} +  U \Psi_{1\uparrow}^\dagger \Psi_{1\uparrow} \Psi_{1\downarrow}^\dagger \Psi_{1\downarrow}.
\end{equation}
Here, the spin is denoted by $\sigma=\uparrow,\downarrow$. The Hamiltonian of the left lead reads
\begin{equation}\label{equ:Ham_lead2}
H_R= \sum_{q \sigma} \omega_q \Psi_{q \sigma}^\dagger \Psi_{q \sigma}.
\end{equation}
Electrons with different spins in lead mode $\vert q  \rangle$ have the same mode energy $\omega_q$. The coupling term between the system and the left lead is given by
\begin{equation}\label{equ:Ham_coup2}
H_{SR}=\sum_q (V_{q \sigma} \Psi_{1\sigma}^\dagger \Psi_q + V_{q \sigma}^* \Psi_q^\dagger \Psi_{1\sigma}).
\end{equation}
$V_{q \sigma}$ is the coupling strength between lead mode $q$ and spin orbital $\sigma$. Even though both Models A and B can accommodate two electrons, there is a qualitative difference between the two. In Model B the lead operators have spin indices so that
\begin{equation} \label{equ:equi1}
\langle \Psi_{q \sigma}^\dagger \Psi_{q' \sigma'} \rangle = n_F(\omega_q-E_F) \delta_{qq'} \delta_{\sigma \sigma'}.
\end{equation}
The additional $\delta_{\sigma \sigma'}$ in Eq.\,(\ref{equ:equi1}) compared to Eq.\,(\ref{equ:equi2}) will be critical, as will be shown later on.

\section{Equations of motion for the reduced system density matrix}\label{sec-master}

Master equations have been widely used to simulate electron transport through quantum systems\cite{brud93,lehm02,li05a,cui05,sven07_1,harb06a}. Using a numerical decomposition of the lead spectral density\cite{meie99,klei04a}, a non-Markovian master equation for electron transport was derived for non-interacting\cite{sven06_1,li07} and interacting\cite{sven06_4} electrons to second-order in $H_{SR}$. Higher order coupling elements can be derived via path-integral calculus\cite{sven07_1}. The approach is valid for arbitrary Coulomb coupling strength, temperature and bias. Master equations can be used to calculate the GF of the charge transfer statistics\cite{max06a,wang07,Bagr03}. 

The total density operator is denoted $\rho$, and $\rho_S=\mathrm{tr}_R \lbrace\rho \rbrace$ and $\rho_R=\mathrm{tr}_S \lbrace \rho \rbrace$ denote the system and the leads components of $\rho$, respectively. We use the full Fock space as a basis for the system part and hereafter we will refer to $\rho_S$ simply as density matrix. The quantum master equation to second-order perturbation theory in $H_{SR}$ reads
\begin{eqnarray} \label{masterorg}
\dot{\rho}_S(t)&=& -i \mathcal L_S \rho_S(t) \\
& &-\mathrm{tr}_R \lbrace \mathcal L_{SR} \int_{t_0}^t \mathrm dt' G_{S+R}(t,t') \mathcal L_{SR} G_{S}^\dagger (t,t') \rho(t) \rbrace \nonumber.
\end{eqnarray}
Eq.\,(\ref{masterorg}) is derived in Appendix \ref{appendix:A}.
The Liouville operators are defined via $\mathcal L_{SR}\cdot=[H_{SR},\cdot]$, $\mathcal L_{S}\cdot=[H_{S},\cdot]$ and the propagators are $G_{S+R}(t,t')=\mathrm{exp}(-i (\mathcal L_S+\mathcal L_R)(t-t'))$ and $G_{S}(t,t')=\mathrm{exp}(-i \mathcal L_S (t-t'))$. We also set $\hbar=1$. In order to propagate Eq.\,(\ref{masterorg}), we derive EOM for its dissipative part (second term in Eq.\,(\ref{masterorg})). This will be done in the following two subsections.

\subsection{Quantum master equation for Model A}\label{sec2-master}

Applying the system-lead coupling term (\ref{equ:Ham_coup}) of Model A to Eq.\,(\ref{masterorg}) one gets\cite{sven06_1,sven06_2}
\begin{eqnarray}\label{equ:mastereigen}
\dot{ \rho}_S(t)&=&-i [H_S, \rho_S(t)]  \nonumber \\ 
& & \big \lbrace -\sum_{s} \big( \Psi_{s}^\dagger \Psi_{s}^{(+)}(t)  \rho_S(t)+\Psi_{s}^\dagger   \rho_S(t) \Psi_{s}^{(-)} (t) \big) \nonumber \\
& & +\sum_{s}\big( \Psi_{s}^{(+)}(t)   \rho_S(t) \Psi_{s}^\dagger  -    \rho_S(t) \Psi_{s}^{(-)} (t) \Psi_{s}^\dagger \big) \nonumber \\
& &+h.c.\big \rbrace
\end{eqnarray}
The auxiliary operators in Eq.\,(\ref{equ:mastereigen}) are given by
\begin{equation}\label{equ:auxeigen1}
\Psi^{(+)}_{s}(t) = \sum_{s'} \int_{t_0}^t \mathrm dt' C^{(+)}_{ss'}(t-t')  G_S(t,t')  \Psi_{s'},
\end{equation}
\begin{equation}\label{equ:auxeigen2}
\Psi^{(-)}_{s}(t) = \sum_{s'} \int_{t_0}^t \mathrm dt' \left( C^{(-)}_{ss'}\left(t-t'\right) \right)^* G_S(t,t')  \Psi_{s'}.
\end{equation}
The correlation functions $C^{(\pm)}_{ss'}(t)$ are discussed in more detail in Appendix (\ref{appendix:B}).
Because we assume that the coupling of the lead to the two dots is symmetric $T_{q1}=T_{q2}$, the correlation functions (Eqs.\,(\ref{equ:bcxx1-nrwa},\ref{equ:bcxx2-nrwa})) which correspond the cross coupling terms, $s\neq s'$, can be written as $C^{(\pm)}_{ss'}(t-t') = C^{(\pm)}_{ss}(t-t')$. Setting $C_{ss'}(t)=0$ for $s \neq s'$ would correspond to the rotating wave approximation (RWA)\cite{harb06a}. 

Liouville-space allows a compact super operator notation. We define the following Liouville space super operators $\mathcal L A=  [H_S,A]$, $\Psi^{(R)} A=A \Psi$, $\Psi^{(L)} A= \Psi A$, $\Psi^{\dagger(R)} A=A \Psi^\dagger$ and $\Psi^{\dagger(L)} A= \Psi^\dagger A$. $A$ is an arbitrary operator in the space of the system. We use $R,L$ to denote left and right super operators and $l,r$ as indices for the left and right lead. Eq.\,(\ref{equ:mastereigen}) finally reads
\begin{equation}\label{equ:liouville:master1}
\dot{ \rho}_S(t)=W_A(t)  \rho_S(t)
\end{equation}
where $W_A(t)$ is
\begin{equation}\label{masterfinal}
W_A(t) \equiv -i \mathcal L - \Pi^l(t) +\Sigma_+^l(t)  + \Sigma_-^l(t).
\end{equation}
The dissipative term is separated into a diagonal contribution
\begin{equation}\label{equ:liouville:master2}
\Pi^l(t)=\sum_{s} \Psi_{s}^{\dagger (L)} \Psi^{(+,L)}_{s}(t) + \Psi_{s}^{\dagger (R)} \Psi_{s}^{(-,R)} (t)+ h.c.
\end{equation}
which leave the number of electrons in the system unchanged and two off-diagonal parts
\begin{equation}\label{equ:liouville:master3}
\Sigma_+^l(t)=\sum_{s} \Psi_{s}^{\dagger (L)} \Psi_{s}^{(-,R)}(t)+\Psi_{s}^{(R)}  \Psi_{s}^{\dagger (-,L)}(t),
\end{equation}
\begin{equation}\label{equ:liouville:master4}
\Sigma_-^l(t)=\sum_{s} \Psi_{s}^{\dagger (R)} \Psi_{s}^{(+,L)}(t) + \Psi_{s}^{ (L)} \Psi_{s}^{\dagger(+,R)}(t),
\end{equation}
which increase or decrease the number of electrons, respectively. $W_A(t)$ is a $16 \times 16$ matrix in Liouville space.
The terms for the right lead $\Pi^r(t)$, $\Sigma_+^r(t)$ and $\Sigma_-^r(t)$ can be derived by replacing $C_{ss'}^{(\pm)}$ in Eqs.\,(\ref{equ:auxeigen1},\ref{equ:auxeigen2}) with the correlation functions of the right lead and have to be added to Eq.\,(\ref{masterfinal}).

In Appendix (\ref{appendix:B}) we discuss the spectral decomposition of the correlation functions into the form $C^{(\pm)}_{ss'}(t)=\sum_{k=1}^{m+m'}  a^{(\pm)}_k e^{\gamma^{(\pm)}_k t}$. Bi-directional counting requires relatively small bias voltages in order to have significant backwards transfer rates. The Fermi function is expanded into Matsubara-frequencies, $i\gamma^{(\pm)}_k$ for $k=m+1,...,m'$, and small bias voltages can be used in our calculations. A numerically efficient way for calculating Eqs.\,(\ref{equ:auxeigen1},\ref{equ:auxeigen2}) is by propagating the EOM\cite{sven06_1}
\begin{eqnarray}\label{equ:diffaux1}
\frac{\partial}{\partial t} \Psi^{(\pm)}_{s,k}(t)&=&  \gamma^{(\pm)}_{k}  \Psi^{(\pm)}_{s,k}(t)-i [H_S, \Psi^{(\pm)}_{s,k}(t)] \nonumber \\
& &+  \sum_{s'} a^{(\pm)}_{k} \Psi_{s'}.
\end{eqnarray}
Summation over the spectral decomposition given by Eq.\,(\ref{specspec}) results in an explicit expressions for the many-body auxiliary $\Psi^{(\pm)}_{s}(t)=\sum_{k=1}^{m+m'} \Psi_{s,k}^{(\pm)}(t)$ operators. Eqs.\,(\ref{equ:liouville:master1} - \ref{equ:diffaux1}) propagated simultaneously in time form a closed set for calculating the reduced density operator $ \rho_S(t)$. An adiabatic switching of the coupling is not required in this method. We are interested in electron counting statistics at steady state which we reach by numerically propagating  $\rho_S^{st}=\rho_S(t\rightarrow\infty)$, $\Pi=\Pi(t\rightarrow\infty) $, $\Sigma_+ = \Sigma_+(t \rightarrow\infty)$ and $\Sigma_-=\Sigma_-(t\rightarrow\infty)$. 

The RWA as applied in Ref.\cite{harb06a} decouples the coherence and population part of the system density matrix in Liouville space eliminating all interferences. For Model A, the RWA applied to Eq.\,(\ref{equ:liouville:master1}) would reduce it to a Pauli rate equation in the eigenbasis of the system. This formal cancellation can be realized as is illustrated by Model B, a single QD. Model B also allows a maximum number of two electrons, but interferences are absent due to the different spin states of the electrons. In the following subsection, we present the Pauli rate equation for Model B without invoking the RWA.

\subsection{Pauli rate equation for Model B}\label{sec1-master}

Inserting the system-lead coupling term (\ref{equ:Ham_wire2}) of Model B into Eq.\,(\ref{equ:Ham_coup2}), we can write
\begin{eqnarray}\label{equ:master2local}
\dot{ \rho}_S(t)&=&-i [H_S,  \rho_S(t)]  \nonumber \\ 
& & \big\lbrace -\sum_{\sigma} \big( \Psi_{1\sigma}^\dagger \Psi_{1\sigma}^{(+)}(t)  \rho_S(t)+\Psi_{1\sigma}^\dagger   \rho_S(t) \Psi_{1\sigma}^{(-)} (t) \big) \nonumber \\
& & +\sum_{\sigma} \big( \Psi_{1\sigma}^{(+)}(t)   \rho_S(t) \Psi_{1\sigma}^\dagger  -    \rho_S(t) \Psi_{1\sigma}^{(-)} (t) \Psi_{1\sigma}^\dagger \big) \nonumber \\
& &+h.c. \big\rbrace
\end{eqnarray}
The auxiliary operators in Eq.\,(\ref{equ:master2local}) are given by
\begin{equation}\label{equ:aux1local}
\Psi^{(+)}_{1\sigma}(t) =  \int_{t_0}^t \mathrm dt' C^{(+)}_{\sigma \sigma}(t-t')  G(t,t')  \Psi_{1\sigma'},
\end{equation}
\begin{equation}\label{equ:aux2local}
\Psi^{(-)}_{1\sigma}(t) =  \int_{t_0}^t \mathrm dt' \left( C^{(-)}_{\sigma \sigma}\left(t-t'\right) \right)^*  G(t,t')  \Psi_{1\sigma'}.
\end{equation}
The correlation functions  $C^{(\pm)}_{\sigma \sigma}(t-t')$ are given in Eqs.\,(\ref{equ:bcxx1},\ref{equ:bcxx2}). The master equation can be recast in the compact Liouville space form 
\begin{equation}\label{equ:liouville:master}
\dot{ \rho}_S(t)=W_B(t)  \rho_S(t)
\end{equation}
with 
\begin{equation}\label{ratecoherence}
W_B(t)\equiv -i \mathcal L - \Pi^l(t) +\Sigma_+^l(t)  + \Sigma_-^l(t).
\end{equation}
The diagonal part is given by
\begin{equation}
\Pi^l(t)=\sum_{\sigma} \Psi_{1\sigma}^{\dagger (L)} \Psi^{(+,L)}_{1\sigma}(t) + \Psi_{1\sigma}^{\dagger (R)} \Psi_{1\sigma}^{(-,R)} (t)+ h.c.
\end{equation}
and off-diagonal parts are
\begin{equation}
\Sigma^l_+(t)=\sum_{\sigma} \Psi_{1\sigma}^{\dagger (L)} \Psi_{1\sigma}^{(-,R)}(t)+\Psi_{1\sigma}^{(R)}  \Psi_{1\sigma}^{\dagger (-,L)}(t),
\end{equation}
\begin{equation}
\Sigma^l_-(t)=\sum_{\sigma} \Psi_{1\sigma}^{\dagger (R)} \Psi_{1\sigma}^{(+,L)}(t) + \Psi_{1\sigma}^{ (L)} \Psi_{1\sigma}^{\dagger(+,R)}(t). 
\end{equation}
Because $C^{(\pm)}_{\uparrow \downarrow}= C^{(\pm)}_{\downarrow \uparrow }=0$ (see Appendix (\ref{appendix:B})) the spin quantum number causes a separation of coherence and population part in the density operator. The population part of Eq.\,(\ref{equ:liouville:master}) satisfies the Pauli rate equation
\begin{equation}
\dot{P}_S(t)=W_P P_S(t).
\end{equation}
$P$ denotes the population of the states and $W_P$ is the Pauli rate matrix.

\section{The generating function}\label{sec3}

The transfer probability of $\bm k=(k_1,k_2)$ electrons in the time interval $t-t_0$ through the left lead is denoted $P(\bm k; t)$. We define the generating function $G(\bm \lambda; t)$ by
\begin{equation}\label{eq:Fourier}
G(i \bm \lambda ; t)= \sum_{\bm k} e^{i \bm k \bm \lambda} P(\bm k; t)
\end{equation}
The probability distribution is obtained using $P(\bm k;t)=\frac{1}{2\pi} \int_0^{2\pi} G(i \bm \lambda;t) \rm e^{-i \bm \lambda \cdot \bm k}$.

We will calculate the GF by tracing the generating operator\cite{max06a,wang07} (GO) $g(\bm \lambda;t)$ over the system degrees of freedom
\begin{equation}
G(\bm \lambda ;t)=  \mathrm{tr}_S \lbrace g(\bm \lambda;t) \rbrace.
\end{equation}
Based on the master equation (\ref{equ:liouville:master}),  $g(\bm \lambda;t)$ satisfies the EOM
\begin{equation}\label{genop}
\dot{g}(\bm \lambda ;t)=W(\bm \lambda) g(\bm \lambda ;t).
\end{equation}
Counting starts after the system has reached steady state. Thus the initial condition of the GO is given by the steady state of the system density matrix $g(\bm \lambda ;t=0)=\rho_S^{st}$. The propagator $W(\bm \lambda)$ of the GO is
\begin{equation}\label{genop1}
W(\bm \lambda)= -i \mathcal L - \Pi^{l}-\Pi^{r}+ e^{\lambda_1} \Sigma_+^l  + e^{\lambda_2} \Sigma_-^l + \Sigma_+^r  + \Sigma_-^r.
\end{equation}
We shall consider the statistics of charge transfers between the left lead and the system. $\bm \lambda:=(\lambda _1, \lambda_2)$ controls the specific statistics obtained by propagating Eq.\,(\ref{genop}). We investigate 3 cases. Setting $\bm \lambda_{+}:=(\lambda _1=\lambda, \lambda_2=0)$ gives the counting statistics of the incoming electrons, $\bm \lambda_{-}:=(\lambda _1=0, \lambda_2=\lambda)$ the outgoing, and $\bm \lambda_{n}:=(\lambda _1=\lambda, \lambda_2=-\lambda)$ the net-process. The corresponding probabilities are denoted by $P^{(\eta)}(k;t)$ with $\eta=+,i,n$, respectively.

\section{Cumulants of the transfer distributions}\label{cumu}

We shall calculate the first $C_1(t)$, the second $C_2(t)$ and the third $C_3(t)$ cumulant of $P^{(\eta})(\bm k;t)$ with respect to $k$. $C_1(t)=\bar k / t=\sum_{\bm k} k P({\bf{k}};t)/t$ is related to the average current 
$
I(t)=e C_1(t).
$
The second cumulant defined by $C_2(t)=(\overline{k^2} - {\bar k}^2)/t$ is commonly represented by the Fano factor $F(t)=\frac{C_2(t)}{C_1(t)}$.
The third cumulant $C_3(t) = \overline{(k-\overline{k})^3}/t$ measures the skewness of the probability distribution
$P(\bm k;t)$ with respect to $k$. For small bias, the current is small and the electrons transferring into the system are uncorrelated. This leads to a Poisson counting statistics where the Fano factor equals $F = 1$. If $F < 1$, the process is Sub-Poissonian, while $F > 1$ indicates Super-Poissonian statistics.

The time-dependent cumulants can be calculated from 
\begin{equation}
C_i^{(\eta)}(t)= \left. \partial_\lambda^i K(\bm \lambda_\eta;t) \right \vert_{\lambda=0}
\end{equation}
where $K(\bm \lambda;t)$ is obtained from the GF
\begin{equation}\label{kkkk}
K(\bm \lambda_\eta;t)=- \frac{1}{t} \mathrm{ln} (G(\bm \lambda_\eta;t)).
\end{equation}
We make use of the fact, that $K(\bm \lambda; t)$ has a well defined long time limit and calculate the asymptotic values\cite{max06a} for the steady state current
\begin{equation}
C_1^{(\eta)}= \left. \frac{\partial}{\partial \lambda} \lim_{t \to \infty} K(\bm \lambda_\eta;t) \right \vert_{\lambda=0}
\end{equation}
and the zero-frequency power spectrum
\begin{equation}
C_2^{(\eta)}=\left. \frac{\partial^2}{\partial \lambda^2} \lim_{t \to \infty} K(\bm \lambda_\eta;t) \right \vert_{\lambda=0}.
\end{equation}
The asymptotic value of the skewness is defined in the same fashion
\begin{equation}
C_3^{(\eta)}= \left. \frac{\partial^3}{\partial \lambda^3} \lim_{t \to \infty} K(\bm \lambda_\eta;t) \right \vert_{\lambda=0}.
\end{equation}
The long time limit of $K(\bm \lambda_\eta;t)$ can be calculated from the dominant eigenvalue\cite{max06} $\epsilon_{1}(\bm \lambda_\eta)$ of the propagator $W(\bm \lambda_\eta)$ given by Eq.\,(\ref{genop1}). Then, $K(\bm \lambda_\eta)= \lim_{t \to \infty} K(\bm \lambda_\eta;t) =\epsilon_{1}(\bm \lambda_\eta)$. Note that $\epsilon_{1}(\bm \lambda_\eta=0)=0$.

\section{Probabilities of elementary events}\label{sec4}

We introduce elementary probabilities to characterize consecutive electron transfer events $m_1$,$m_2$ at times $t_1,t_2$. $m$ characterizes the side of the process $(l,r)$ and if the electron is transferred in (+) or out (-) of the QD. The elementary probability is given by\cite{gardiner,Muka03a}
\begin{equation}\label{prob2}
P(t_2, t_1) = \langle \Sigma^{m_2} S_{t_2,t_{1}} \Sigma^{m_1} S_{t_1,t_{0}} \rangle.
\end{equation}
Eq.\,(\ref{prob2}) is the joint probability of detecting specified electron transfers at times $t_1$ and $t_2$ with no transfers occurring in the time intervals $t_1,t_0$ and  $t_2,t_1$.
We denote the trace over the system degrees of freedom by $\langle \cdot \rangle=\mathrm{tr}_S\lbrace \cdot \rho_S(t_0) \rbrace$. $S_{t_i,t_j}$ is the propagator of the system in absence of transfer events at the leads within the time interval $t_i, t_j$:
\begin{equation}\label{prob2-S}
S_{t_i,t_{j}} = \mathrm{exp} \left( \left(-i \mathcal{L}-\Pi^{l}-\Pi^{r} \right) \left(t_i-t_{j}\right) \right).
\end{equation}

We shall consider three cases. $i)$ An electron is detected when it enters the junction through the left lead at time $t_1 = 0$ and leaves the junction through the right lead at time $t$. The electron transfer operators are $\Sigma^l_1$ and $\Sigma^r_2$ respectively and we denote this transfer pathway by $l \rightarrow r$. The joint elementary probability is then given by
\begin{equation}\label{lr1}
P_{l \rightarrow r}(t, t_0) = \langle \Sigma^r_-  S_{t,t_{0}} \Sigma^l_+ \rangle.
\end{equation}
$ii)$ The reverse process, $r \rightarrow l$, which we write as
\begin{equation}\label{lr2}
P_{r \rightarrow l}(t, t_0) = \langle \Sigma^l_-  S_{t,t_{0}} \Sigma^r_+ \rangle.
\end{equation}
$iii)$ Transfer from the left lead into the system and back into the left lead denoted by
$l \rightarrow l$ can be written as
\begin{equation}\label{lr3}
P_{l \rightarrow l}(t, t_0) = \langle \Sigma^l_-  S_{t,t_{0}} \Sigma^l_+ \rangle.
\end{equation}

Quantities (\ref{lr1}-\ref{lr3}) can be measured as follows. One has to detect single directionally resolved electron transfers between the leads and the system and record a sufficiently long time-series of transfer events. Then a histogram of the number of occurrences of a specific consecutive transfer event, such as $l \rightarrow r$, as function of increasing time intervals ${t-t_0}$ can be generated. The histogram has to be normalized by the total number of events $\Sigma^{m_1} \rightarrow \Sigma^{m_2}$ in the time series.

The total conditional probability $P^c_{m_1 \rightarrow m_2}$ to have an $\Sigma^{m_2}$ event if the last event was a $\Sigma^{m_1}$ event irrespective of the time-interval between $m_2$ and $m_1$ can be calculated by integrating the time-dependent conditional probability $P^c_{m_1 \rightarrow m_2}(\tau \vert t_0)= \langle \Sigma^{m_2}  S_{\tau,t_0} \Sigma^{m_1} \rangle/ \langle \Sigma^{m_1} \rangle$:
\begin{equation}\label{probint}
P^c_{m_1 \rightarrow m_2}=\int_0^\infty \mathrm d\tau \langle \Sigma^{m_2}  S_{\tau,t_0} \Sigma^{m_1} \rangle/ \langle \Sigma^{m_1} \rangle.
\end{equation}
Other interesting quantities can be calculated from Eqs.\,(\ref{lr1}-\ref{lr3}). For example, the residence time of electrons in the system subject to a specific transfer process $m_1\rightarrow m_2$ is given by
\begin{equation}\label{Eq:wait-av}
t_{res}=\frac{\int_{t_0}^\infty \mathrm d\tau \, \tau \langle \Sigma^{m_2} S_{\tau,t_0} \Sigma^{m_1} \rangle}{\int_0^\infty \mathrm d\tau \langle \Sigma^{m_2} S_{\tau,t_0} \Sigma^{m_1} \rangle}.
\end{equation}
We propose another interesting setup by counting electrons only at the left lead regardless of electron transfers occurring at the right lead. Thus the propagator (\ref{prob2-S}) is modified to 
\begin{equation}
\tilde S_{t_i,t_{j}} = \mathrm{exp} \left( \left(-i \mathcal{L}-\Pi^{l}-\Pi^{r}+ \Sigma_+^r  + \Sigma_-^r \right) \left(t_i-t_{j}\right) \right).
\end{equation}
The conditional probability of an electron entering the system at time $t_0$ through the left lead, and the next electron entering at time $t$ from the left lead can be written as
\begin{equation}\label{lr4}
P^{c}_{l,l}(t \vert t_0) = \langle \Sigma^l_+  \tilde S_{t,t_0} \Sigma^l_+ \rangle/ \langle \Sigma^l_+ \rangle.
\end{equation}
We denote this transfer series by $l,l$. In the time interval $t$-$t_0$ electron transfers take place only at the right lead. For uncorrelated electron transfer, the propability distribution (\ref{lr4}) is Poissonian and can be written as\cite{bronstein}
\begin{equation}\label{lr5}
P^{poiss}_{l,l}(t \vert t_0) = e^{-C_1 t}\langle\Sigma^l_+\rangle.
\end{equation}

\section{Numerical Simulations}\label{sec5}

The master equation for Model A (\ref{equ:liouville:master1}) and the EOM of the auxiliary creation operator (\ref{equ:diffaux1}) form a system of equations of motion and we propagate them simultaneously into steady state using the Runge-Kutta method. At steady state, the memory of the non-Markovian master equation vanishes and Eq.\,(\ref{equ:liouville:master1}) corresponds to its Markovian counterpart and the electron transfer operators Eqs.\,(\ref{equ:liouville:master2}-\ref{equ:liouville:master4}) become time-independent. Using the obtained transfer operators and steady state density matrix, we then propagate the EOM of the generating operator (\ref{genop}) to a finite binning time $t$ for each step of the discretized counting field $\lambda$ to derive the time-dependent cumulants. Alternatively, we use an eigenvalue decomposition of the steady state propagator of Eq.\,(\ref{genop1}) in order to calculate the asymptotic cumulants. The same approach is used for Model B based on Eq.\,(\ref{equ:liouville:master}) and its dependencies.

\begin{figure}
\includegraphics[width=5cm,clip]{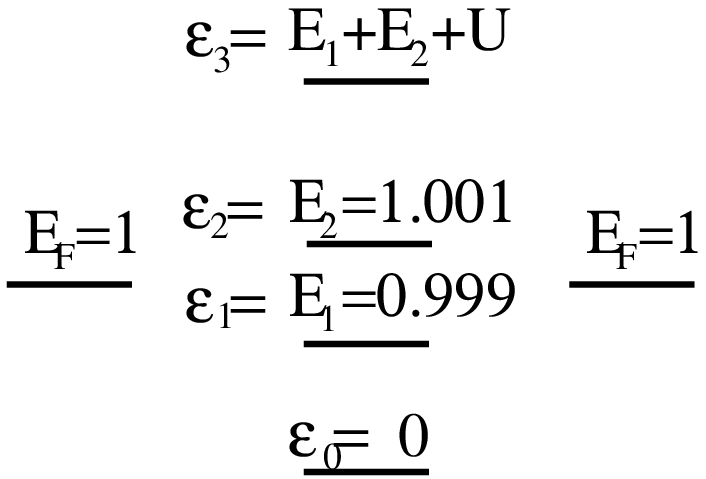}
\caption{Level scheme in many-body eigenbasis for Model A. For Model B, $E_1$ and $E_2$ have to be replaced by $E_{1\uparrow}$ and $E_{1\downarrow}$.
\label{fig:88}}
\end{figure}

We shall expand all system operators in the many-body eigenbasis of the system. For Model A, we use the transformation
\begin{equation}
\Psi^\dagger_s =\sum_{mm'} \alpha_{mm'}^{(s)} \vert m \rangle \langle m' \vert.
\end{equation}
The many body basis is spanned by four states $\vert 0\rangle =\vert 0 0\rangle$, $\vert 1\rangle =\vert 0 1\rangle$, $\vert 2\rangle =\vert 1 0\rangle$ and $\vert 3\rangle =\vert 1 1\rangle$. The coefficients of the transformation $\alpha_{mm'}^{(\dagger,s)}$ between orbital basis and many-body basis can be derived from the Fermion anticommutator relations. Here the non-zero coefficients for the creation operators are $\alpha_{20}^{(1)}=1$, $\alpha_{31}^{(1)}=1$. $\alpha_{10}^{(2)}=1$ and $\alpha_{32}^{(2)}=-1$. Thus we get $\Psi^\dagger_1=\vert 1 0 \rangle \langle 0 0\vert +\vert 1 1 \rangle \langle 0 1\vert$ and $\Psi^\dagger_2=\vert 0 1 \rangle \langle 0 0\vert - \vert 1 1 \rangle \langle 1 0\vert$. The annihilation operators can be derived by replacing $\alpha_{mm'}^{(s)}$ with $\alpha_{m'm}^{(s)}$. For Model B, one has to replace $s=1$ with $\sigma=\uparrow$ and $s=2$ with $\sigma=\downarrow$. 
We can write (\ref{equ:Ham_wire}) and (\ref{equ:Ham_wire2}) as 
\begin{equation}
H_S=\sum_{m=0}^3 \epsilon_m \vert m\rangle \langle m \vert.
\end{equation}
$\epsilon_m$ is the energy of state $\vert m\rangle$. The energy scheme we used for Model A and B is shown in Fig.\,(\ref{fig:88}). All energies are scaled with respect to the equilibrium chemical potential of the leads $E_F=1$. The fixed orbital energies for Model A are $\epsilon_1=E_1=0.999$, $\epsilon_2=E_2=1.001$ and for Model B $\epsilon_1=E_{1\uparrow}=0.999$ and $\epsilon_1=E_{1\downarrow}=1.001$. Thus, the double occupancy state has an energy of $\epsilon_3=E_1+E_2+U=2.0+U$. The ground state is set to zero $\epsilon_0=0$. The bias voltage is applied symmetrically to the system using $E_{F,l}=E_F+V/2$, $E_{F,r}=E_F-V/2$. To study temperature fluctuations of electron transfer as a function of $U$, we set $T=0.002$ to be in a temperature range of $\beta V\approx U$. In Eq.\,(\ref{equ:spectralnum}), we restrict the spectrum to a single Lorentzian centered at $\Omega_1=E_F$. We also chose relatively large bandwith parameter $\Gamma_1=1$ (wide-band limit). We set $p_1=2\cdot 10^{-4}$ in Eq.\,(\ref{equ:spectralnum}) ($p_1\sim \sum_q \vert V_q\vert^2 = \sum_q \vert T_q\vert^2$) as a reasonable small value considering the weak coupling required in order to guarantee physical results within second-order perturbation theory.

\subsection{The bidirectional transfer probability}

\begin{figure}
\includegraphics[width=8.5cm,clip]{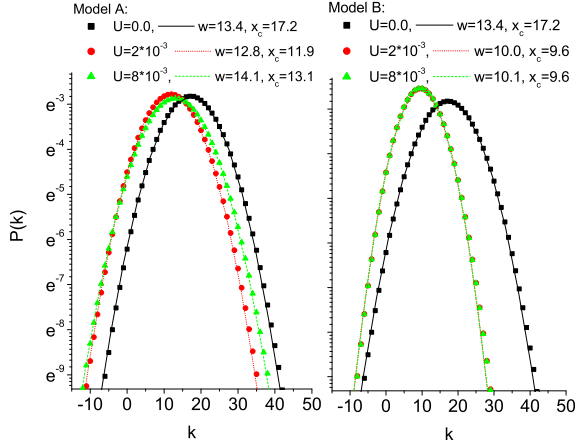}
\caption{Electron transfer probability $P^{n}(k,t)$ for different Coulomb coupling strengths $U$ as function of the net-number of transferred electrons $k$. Left panel depicts Model A, right frame Model B. The time is set to $t = 200$. A small bias is applied $V=4\cdot 10^{-4}$. $w:=$\textit{variance} and $x_c:=$\textit{position of peak} are the fit parameters of the Gaussian. In Model B, the curves for $U=2\cdot 10^{-3}$ and $U=8\cdot 10^{-3}$ overlap.
\label{fig:6}}
\end{figure}

In Fig.\,\ref{fig:6} we show the probability $P^{(n)}(k;t)$ of the net-number $k=k_1-k_2$ of electron transfers between the left lead and the system at steady state. The binning time is fixed to $t = 200$. The left panel depicts Model A, the right panel Model B. The distributions for different Coulomb coupling strengths $U = 0$ , $U = 2\cdot 10^{-3}$ and $U = 8\cdot 10^{-3}$ are fitted with a Gaussian $P(k)=\frac{\sqrt{w}}{2\sqrt{2}}\mathrm{exp}(-\frac{2}{w}(k-x_c)^2)$. The fit parameters are given in the graph. A small bias of $V = 4\cdot 10^{-4}$ allows electron transfers against the bias. As discussed for non-interacting\cite{max06a} and strongly coupled\cite{Bagr03} electrons, the Gaussian provides a good approximation for $P^{(n)}(k;t)$. This is mainly due to small non-equilibrium contrains. For high bias, the deviations will be more significant. Interference effects are enhanced by Coulomb interaction. This can be seen by comparing the results for Models A and B in Fig.\,\ref{fig:6} where interferences are present and absent, respectively. For $U=0.0$, the probability distributions for both models are the same. With increasing $U$, the fitted curve is shifted and broadended in the presence of interferences.

\subsection{The fluctuation theorem}

\begin{figure}
\includegraphics[width=8.5cm,clip]{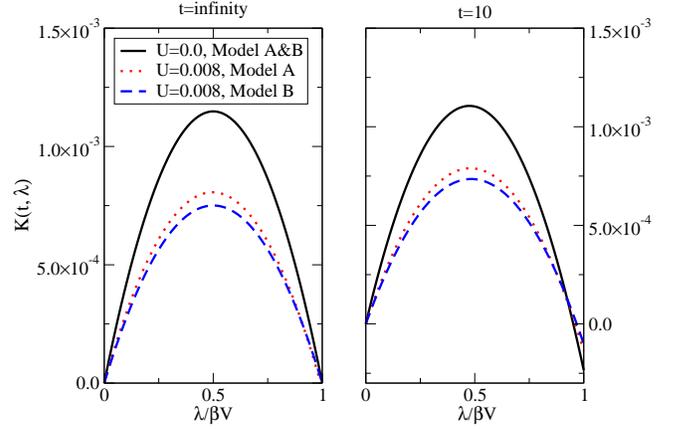}
\caption{Cumulant generating function $K(t,\bm \lambda_{n})$ of the net-process as a function of $\lambda$ for different values of Coulomb repulsion $U$. Left panel shows results obtained for infinite binning time $\lim_{t\rightarrow \infty} K(t,\bm \lambda_{n})$, right panel for a finite binning time of $t=10$. The x-axis is rescaled by $\beta V$.
\label{fig:13}}
\end{figure}

Fig.\,\ref{fig:13} shows $K(t,\bm \lambda_{n})$, given in Eq.\,(\ref{kkkk}), for the net process as a function of $\lambda$ for different Coulomb coupling strengths. The x-axis was rescaled by the fixed bias of $V=2\cdot 10^{-3}$. To explore the role of interferences we compare Models A and B. The left panel shows the asymptotic cumulant generating function $\lim_{t \to \infty} K(t,\bm \lambda_{n})$ which is computed from the eigenvalues of $W(\bm \lambda_{n})$, the right panel shows $K(t,\bm \lambda_{n})$ for finite binning time $t=10$ calculated by propagation the EOM (\ref{genop}). We observe that finite Coulomb coupling, $U=2\cdot 10^{-3}$, significantly changes the cumulant GF $K(t,\bm \lambda_{n})$. This indicates that the influence of Coulomb blockade can be measured in the first cumulant, the current. The effects due to the interference terms are smaller and are observable for $U\neq 0$. In this regime, a discrepancy between Model A and B can be observed for $\lambda/\beta V \approx 0.5$ which indicates that it is significant in the higher cumulants. We find that at infinite binning time the symmetry $K(t,\bm \lambda)$=$K(t,\bm \lambda-\beta V)$ holds in both models as shown in the left panel of Fig.\,\ref{fig:13}. Immediately follows from this symmetry the fluctuation theorem, $\frac{P(k)}{P(-k)}=\mathrm{exp}(\beta V k)$, for $t \rightarrow \infty$. This relation thus holds in systems with interferences and Coulomb interaction.

\subsection{Cumulants of the transfer probability distribution}

In the following, we discuss the effects of quantum interference and Coulomb coupling on first cumulant $C_1$, the Fano factor $F=C_2/C_1$¸ and the normalized skewness $C_3/C_1$ for a infinite binning time. 

\begin{figure}
\includegraphics[width=8.5cm,clip]{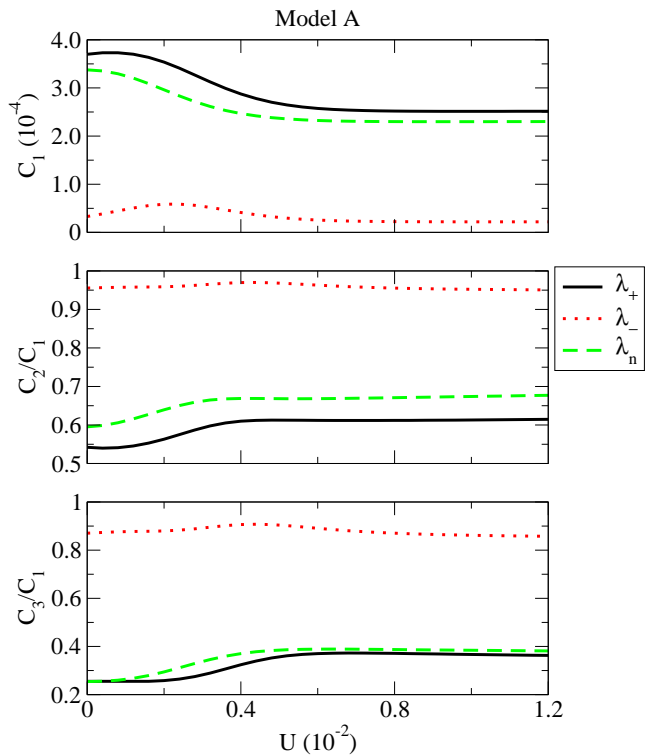}
\caption{First cumulant $C_1$ (top panel), Fano factor $C_2/C_1$ (middle panel)¸ and the normalized skewness $C_3/C_1$ (bottom panel) as functions of Coulomb coupling $U$ for infinite binning time. $\bm \lambda_+$ refers to the incoming, ¸$\bm \lambda_-$ the outgoing, and $\bm \lambda_{n}$ the net-transfers of electrons between the left lead and the system. The bias voltage is $V=1\cdot 10^{-3}$.
\label{fig:8}}
\end{figure}

The upper panel of Fig.\,\ref{fig:8} depicts the first cumulant through the left lead for three different processes: $\bm \lambda_+$ (ingoing), $\bm \lambda_-$ (outgoing) and $\bm \lambda_{n}$ (net-process). The bias is $V = 1\cdot 10^{-3}$. $U$ introduces an energetic penalty for double occupancy. This explains the current drop around $U = 2\cdot 10^{-3}$ by the fact that the energy of the double occupancy state increases with $U$. This reduces the occupancy of the state and its contribution to the total average current. The drop is smooth due to the finite temperature. Interesting effects can be observed at the intermediate coupling regime. So is the drop of the net-current accompanied with an increase of the reverse current. The Fano factor and the normalized skewness of the transfer in direction of the bias as well as the net-transfer show an increase following the decrease in current due to increasing $U$. The corresponding values for the reverse transfer show a weak response.

\begin{figure}
\includegraphics[width=8.5cm,clip]{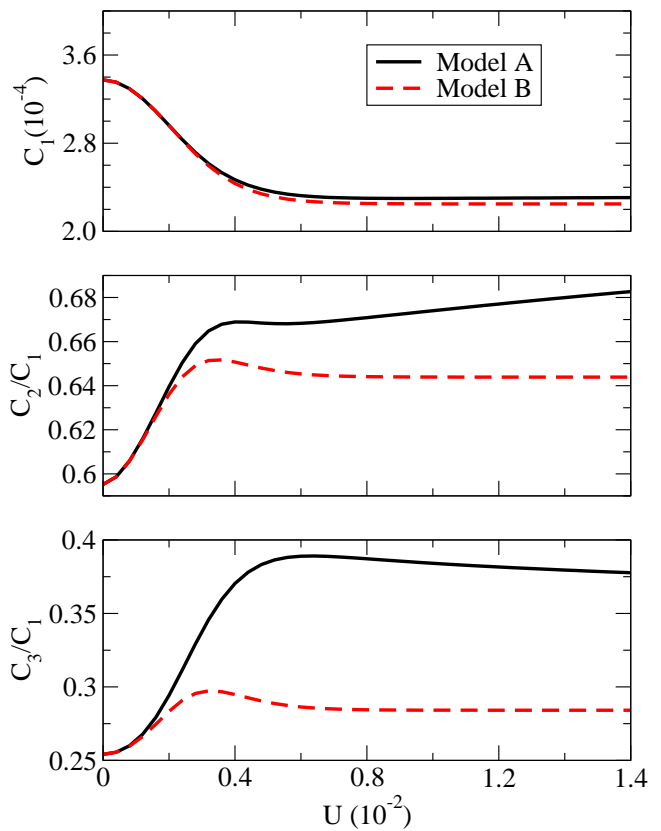}
\caption{Comparison of first cumulant, Fano factor and normalized skewness for the net-transfer statistics ($\bm \lambda_{n}$) between Model A and B for infinite binning time. The bias is $V=1\cdot 10^{-3}$.
\label{fig:7}}
\end{figure}

Comparison of Model A with Model B shown in Fig.\,\ref{fig:7} demonstrates the influence of interferences on the net-transfer statistics ($\bm \lambda_{n}$). The skewness is of particular interest since it was found to be the most sensitive of the three in systems with interference\cite{wang07}. The Fano factors of Model A and B diverge in Fig.\,\ref{fig:7} with increasing $U$ but their absolute difference remains small. The skewness reveals the presence of interferences more clearly than the Fano factor. The average current shows a negligible difference. The observability of interferences requires a strong Coulomb coupling since the differences between Model A and B in the third cumulants are significant when the double occupancy state is inaccessible by electronic excitations.

\subsection{Joint elementary probabilities}

\begin{figure}
\includegraphics[width=8.5cm,clip]{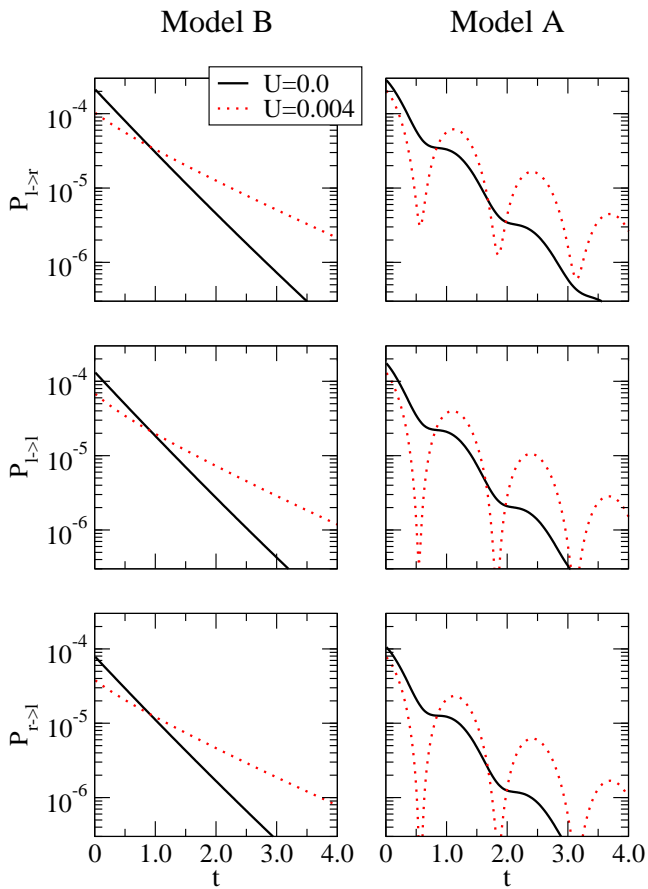}
\caption{Elementary probabilities $P_{l \rightarrow r}(t, t_0)$/$P_{l \rightarrow l}(t, t_0)$/
$P_{r \rightarrow l}(t, t_0)$ for an electron entering the junction through the
left/left/right lead at time $t_0$ and leaving through the right/left/left lead at time $t$, respectively. The left side shows Model B, the right side Model A. The bias voltage is $V =2\cdot 10^{-3}$.
\label{fig:3}}
\end{figure}

Using expressions (\ref{lr1}-\ref{lr3}), we calculated the joint probabilities of directionally resolved consecutive electron transfers. Fig.\,\ref{fig:3} compares the joint probabilities $P_{l\rightarrow r}(t, t_0)$, $P_{l\rightarrow l}(t, t_0)$, $P_{r\rightarrow l}(t, t_0)$ for the two models and a small bias of
$V = 2\cdot 10^{-3}$. In Model B, we observe an almost exponential decay of the probability of an outgoing electron transfer event at the left/left/right site following an incoming electron
transfer event at $t_0$ at the right/left/left site. A weak non-exponential slope indicates a weak correlation between the transfer processes. The probabilities of the reverse processes $r \rightarrow l$ decay faster than in the direction of the bias $l \rightarrow r$. Consequently, the $l \rightarrow l$ process has an intermediate decay rate. Electron-electron coupling leads to a slower decay in all three processes, indicating that the probability of an electron to reside longer within the junction increases with $U$. The probability has its maximum at $t_0$ since the orbitals are in direct contact with both leads. 

\begin{figure}
\includegraphics[width=8.5cm,clip]{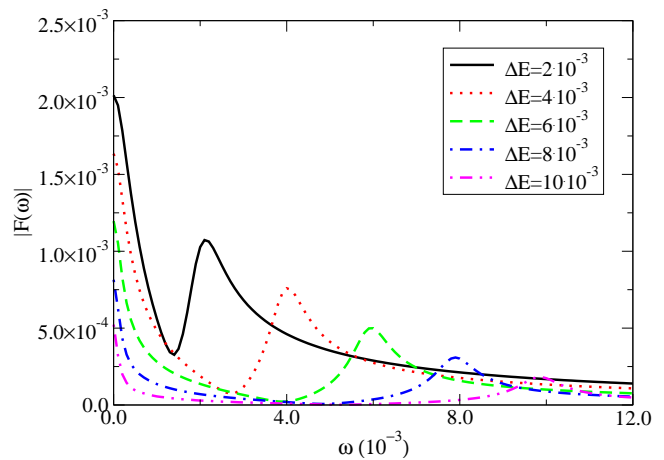}
\caption{Magnitude $\vert F(\omega)\vert$ of the Fourier transformation of $P_{l \rightarrow r}(t, t_0)$ for Model A. $\Delta E=E_1-E_2$ is the energy difference between orbits of the quantum dots. Coulomb coupling is set to $U=4\cdot 10^{-3}$. The bias voltage is $V =2\cdot 10^{-3}$.
\label{fig:3f}}
\end{figure}

In Model A, the exponential decay of the probability is superimposed by an oscillation due to orbital interference. The amplitude of the oscillation is increased by Coulomb interaction. We Fourier transformed the $P_{l \rightarrow r}(t, t_0)$ probability of Model A in order to analyze the dependency of the frequency on the system parameters. The magnitude of $F(\omega)=\frac{1}{\sqrt{2 \pi}} \int_{0}^\infty \mathrm dt e^{-i \omega t} P_{l \rightarrow r}(t,t_0=0)$ is shown in Fig.\,\ref{fig:3f}. From the peaks we can conclude that the frequency $\omega_{l \rightarrow r}$ of the oscillation in $P_{l \rightarrow r}(t,t_0=0)$ is determined by the detuning of the orbitals of the two quantum dots $\omega_{l \rightarrow r}=E_1-E_2=\epsilon_1-\epsilon_2$. We could not find a dependence of $\omega_{l \rightarrow r}$ on $U$. The energy gap between the double occupancy state and the single occupancy states $\epsilon_3-\epsilon_{1}$ ($\epsilon_3-\epsilon_{2}$) is three orders of magnitude larger than the gap between the single occupancy states. We also found that the amplitude of the oscillation decreases with increasing detuning. Thus oscillations due to the $\epsilon_3-\epsilon_{1}$ ($\epsilon_3-\epsilon_{2}$) energy gap which would depend on $U$ have a high freqency and a small amplitude invisible to our numerical method. The time-series analysis of consecutive electron transfer events seems to be highly sensitive to very small energy differences when interferences are present. Note that this method works for energy differences smaller than $k_b T$, here $\epsilon_1-\epsilon_2 \approx k_b T$, which cannot be resolved by average current measurements by scanning the voltage of the leads.

\begin{figure}
\includegraphics[width=8.5cm,clip]{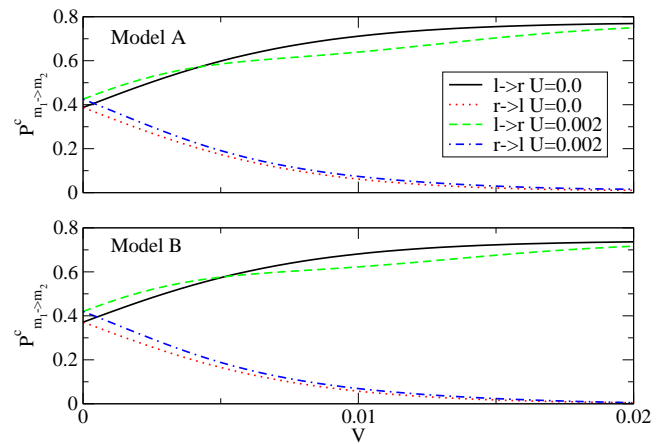}
\caption{Total conditional probabilities (Eq.\ref{probint}) $P^c_{l \rightarrow r}$ and $P^c_{r \rightarrow l}$ as a function of bias voltage for different Coulomb coupling strength $U$. Upper panel Model A, bottom panel Model B.\label{fig:4}}
\end{figure}

The upper panel of Fig.\,\ref{fig:4} depicts the total conditional probability $P^c_{m_1 \rightarrow m_2}$ as a function of external bias voltage for Model A (bottom panel Model B). The two transfer processes under consideration are $l \rightarrow r$, $r \rightarrow l$, and we use $U=0.0$ and $U=2\cdot 10^{-3}$ for comparison. Since the probability in Eq.\,(\ref{probint}) is conditional on the first transfer, the $r \rightarrow l$ and $l \rightarrow l$ processes are equally likely. In the absence of an external bias, $P^c_{l\rightarrow r}$ and $P^c_{r\rightarrow l}$ are equal and the electron transfers are driven by the thermal fluctuations of the leads at finite temperature. A splitting occurs when the voltage is turned on. The deviations caused by the Coulomb coupling vanish at high bias in both models.

\begin{figure}
\includegraphics[width=8.5cm,clip]{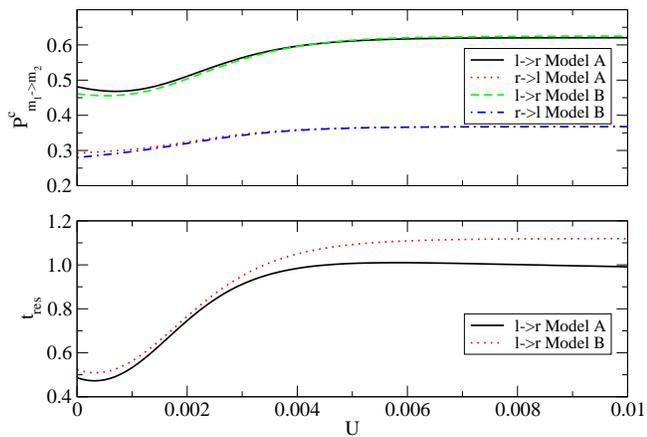}
\caption{
Upper panel: Time-integrated conditional joint probabilities (Eq.\ref{probint}), $P^c_{l \rightarrow r}$ and $P^c_{r \rightarrow l}$ versus Coulomb coupling strength $U$. 
Bottom panel: The mean residence time $t_{res}$ of an electron in the system as a function of $U$ for the different processes and Model A and B. The bias is $V=1\cdot 10^{-3}$
\label{fig:5}}
\end{figure}

Fig.\,\ref{fig:5} uses the same parameters as Fig.\,\ref{fig:4}, but the observables are shown as functions of the Coulomb coupling $U$ at a fixed bias of $V = 2\cdot 10^{-3}$. The upper panel depicts the total conditional probabilities, the bottom panel the average residence times for Model A and B. Increasing $U$ causes a dip of the $l \rightarrow r$ probabilities around $U = 2\cdot 10^{-3}$. There are no qualitative differences between the probabilities of the two models. $U$ strongly affects the mean residence time of the electrons which is also sensitive to the presents of orbital interferences. Thus the residence time between the electron entering and leaving the junction provides a useful measure for the presence of Coulomb coupling and interferences.

\begin{figure}
\includegraphics[width=8.5cm,clip]{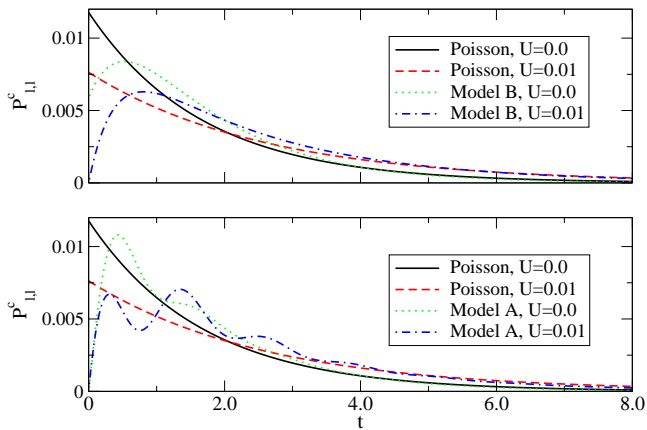}
\caption{Conditional probabilities $P^{c}_{l,l}(t \vert t_0)$ for an electron entering the junction through the left lead at time $t_0$ and the next electron enters at $t$. Only transfers through the left lead are detected. Electrons are allowed to leave and enter through the right lead at all times. Model A and B are compared for different $U$. For comparison, we also plot the corresponding Poissonian processes $P^{poiss}_{l,l}(t \vert t_0)$. The bias voltage is $V= 2\cdot 10^{-3}$.
\label{fig:3a}}
\end{figure}

Fig.\,\ref{fig:3a} depicts the conditional joint probability $P^{c}_{l,l}(t \vert t_0)$, Eq.\,(\ref{lr4}), for an electron to enter the junction through the left lead at time $t_0$ and next electron to enter at time $t$ also through the left lead. The detector is only applied to the left lead, transfers at the right lead are permitted at all times. Models A and B are compared for different $U$. The probabilities of Poissonian two-electron transfers, Eq.\,(\ref{lr5}), are shown for comparison. Similar to the configuration with electron detectors at both leads, Fig.\,\ref{fig:3}, we observe longer tails in the probability distribution in time for larger $U$. Consecutive electron transfers are strongly correlated at short times leading to small probabilities for the next electron to enter the junction after the first one has entered. At intermediate times, the probability is larger than the probability of the Poissonian process. For long times, the transfer becomes weakly correlated and the probabilities for Model A and B are close to a Poissonian distribution. The oscillations are present due to orbital interference in Model A only.

\section{Conclusions}\label{sec6}

We have calculated the counting statistics in a model junction for finite Coulomb coupling strength, bias and finite temperatures. The numerical results reveal several significant measurable effects of the quantum interference in a DQD on the electron transfer statistics. 

The skewness provides a sensitive observable. We also show that a measurement of the average residence time of electrons is affected by quantum mechanical interference as well as the Coulomb coupling between two parallel quantum dots. The observed oscillations in the joint elementary probabilities can be recovered by time-series analysis. Their amplitude and frequency are directly related to the Coulomb coupling and the level detuning of the DQD, respectively. 

Several extensions of the model could be of interest. Decoherence effects can be included by coupling a dissipative phonon bath to the sites. Including higher order coupling elements beyond second-order perturbation theory in the GF could reveal additional insights into the dynamics of the system-lead contact.

\acknowledgments 
The support of the National Science Foundation (Grant No. CHE-0446555, CBC-0533162) and NIRT
(Grant No. EEC 0303389)) is gratefully acknowledged. M.E. was partially supported by the FNRS Belgium (charg\'{e} de recherche)

\appendix

\section{Derivation of the quantum master equation} \label{appendix:A}

In this section, we present the derivation of the Nakajima-Zwanzig operator identity\cite{naka58,zwan61,zwan64} and utilize it to couple an lead to the relevant system, here the quantum dots. We define a projection operator $P$, with $P^2=P$, which acts on an arbitrary operator $A$ defined in the Hilbert space of the full system 
\begin{equation}
P A = \mathrm B \, \mathrm{tr}_R \lbrace A \rbrace.
\end{equation}
$B$ is an operator defined in the lead part only. Applying the projection operator to the density operator of the full system leads to
\begin{equation}\label{equ:sepa}
P \rho(t)=\rho_R \, \mathrm{tr}_R\lbrace\rho(t) \rbrace = \rho_R \otimes \rho_S(t).
\end{equation}
Thus, the EOM for the full system can be written as
\begin{equation}
\dot{\rho}(t) = -i \mathcal L \rho(t) =-i \mathcal L P \rho(t) - i \mathcal L Q \rho(t)
\end{equation}
which can be interpreted as the evolution of the projected part $P {\rho}(t)$ plus the evolution of its orthogonal complement $Q\rho=(1-P)\rho$.
Applying the projection operator to the Liouville equation and to its orthogonal complement leads to
\begin{equation}\label{equ:relevant}
P\dot{\rho}(t)= -i P \mathcal L P \rho(t) - i P \mathcal L Q \rho(t)
\end{equation}
and
\begin{equation}\label{equ:compl}
Q \dot{\rho}(t) = -i Q \mathcal{L} P \rho(t) - i Q \mathcal L Q \rho(t),
\end{equation}
respectively. Integrating the differential equation of the orthogonal complement part (\ref{equ:compl}) and applying it to Eq. (\ref{equ:relevant}) results in the Nakajima-Zwanzig operator identity
\begin{eqnarray}
P\dot{\rho}&=&-i P \mathcal L P \rho(t)  \\
& &-i P \mathcal L \vec T e^{-i \int_{t_0}^t \mathrm d \tau (1-P) \mathcal L}
(1-P) \rho(t_0) \nonumber \\  
& &+P \mathcal L \int_{t_0}^t \mathrm d t' \, \vec T e^{-i \int_{t'}^{t} \mathrm d \tau (1-P) \mathcal L}
(1-P) \mathcal L P \rho(t') \nonumber
\end{eqnarray}
which is valid for arbitrary time-dependent Hamiltonians. $\vec T$ is the positive time-ordering operator. One can further simplify by tracing over the lead degrees of freedom and employing the property of the projection operator $\mathrm{tr}_R \lbrace P \dot{\rho}(t) \rbrace = \dot \rho_S(t)$. This gives
\begin{eqnarray}\label{equ:NZ-kernel}
\dot{\rho}_S(t)&=&-i \mathcal L_S \rho_S(t) -i \mathrm{tr}_R \lbrace \mathcal L_{SR} \rho_R \rbrace \nonumber \\
& &+\int_{t_0}^t \mathrm dt' K(t,t') \rho_S(t') + \mathrm{In}(t)
\end{eqnarray}
where the Kernel $K(t,t')$ reads
\begin{eqnarray}
K(t,t')&=&-\mathrm{tr}_R \lbrace \mathcal L_{SR} \vec T e^{-i \int_{t'}^t \mathrm d \tau (1-P) \mathcal L} \nonumber \\
& &\times (1-P) ( \mathcal L_R +\mathcal L_{SR} ) \rho_R \rbrace
\end{eqnarray}
and the initial value term is given by
\begin{equation}\label{equ:NZ-initial}
\mathrm{In}(t)=-i \mathrm{tr}_R \lbrace \mathcal L_S \vec T e^{-i \int_{t_0}^t \mathrm d \tau (1-P) \mathcal L}(1-P) \rho(t_0) \rbrace.
\end{equation}
In order to derive a practical method for solving Eq.\ (\ref{equ:NZ-kernel}), we utilize second order perturbation theory by assuming that $(1-P) \mathcal L_{SR}\approx 0$ or $\mathcal L_{SR}\approx P\mathcal L_{SR}$, what implies for the Kernel $K(t,t')$ that
\begin{equation}\label{equ:pertu}
e^{-i(1-P) \mathcal L t}=e^{-i(1-P) (\mathcal L_S+\mathcal L_R+\mathcal L_{SR})t} \approx e^{-i(1-P) (\mathcal L_S +\mathcal L_R) t}.
\end{equation}
We can further simplify Eq. (\ref{equ:NZ-kernel}) by making use of the following exact relations:
$
\mathcal L_R \rho_R =0,
$
$
P \mathcal L_{SR} \rho_R=0,
$
$
e^{-i(1-P) \mathcal Lt}=P+(1-P) e^{-i\mathcal L t},
$
$
\mathrm{tr}_R \lbrace \mathcal L_{SR} P C \rbrace=0 
$
for an arbitrary operator C. Finally, neglecting the initial value term (\ref{equ:NZ-initial}) in the evolution of the reduced density matrix results in the Liouville equation with the system-lead coupling term in a time-nonlocal (TNL) regime
\begin{eqnarray} \label{equ:NZtraced}
\dot{\rho}_S(t)&=& -i \mathcal L_S \rho_S(t) \\
& &-\mathrm{tr}_R \lbrace \mathcal L_{SR} \int_{t_0}^t \mathrm dt' G_{S+R}(t,t') \mathcal L_{SR} \rho(t') \rbrace \nonumber
\end{eqnarray}
We neglect the influence of dissipation during the propagation of the density operator by applying the substitution\cite{klei04a}
\begin{equation}
\rho(t')=G_S^\dagger(t,t')\rho(t)
\end{equation}
to the TNL Kernel. This leads to the time-local (TL) description given by Eq.\,(\ref{masterorg}). For the given system, both TNL and TL description produce similar results in the weak coupling regime\cite{sven06_2}.

\section{Numerical decomposition of the spectral density}  \label{appendix:B}

In Model B, the trace over the lead degrees of freedom in the dissipation term of Eq.\,(\ref{masterorg}) can be recast in terms of correlation functions of the form
\begin{equation} \label{equ:bcxx1}
C^{(+)}_{\sigma \sigma'}(t)=\sum_{q}  V_{q\sigma} V_{q\sigma'}^*  \langle  \Psi_{q \sigma'} e^{-i H_R t} \Psi_{q \sigma}^\dagger e^{i H_R t} \rho_R \rangle_R
\end{equation}
\begin{equation} \label{equ:bcxx2}
C^{(-)}_{\sigma \sigma'}(t)=\sum_{q} V_{q\sigma} V_{q\sigma'}^* \langle e^{-i H_R t}  \Psi_{q \sigma}^\dagger  e^{i H_R t} \Psi_{q \sigma'}  \rho_R \rangle_R
\end{equation}
The properties of the trace lead to $C^{(\pm)}_{\uparrow \downarrow}= C^{(\pm)}_{\downarrow \uparrow }=0$. While not an approximation, the procedure mimics a rotating wave approximation in Eq.\,(\ref{equ:liouville:master}).
The correlation functions in Model A used in Eqs.\,(\ref{equ:auxeigen1},\ref{equ:auxeigen2}) are given by
\begin{equation} \label{equ:bcxx1-nrwa}
C^{(+)}_{ss'}(t)=\sum_{q}  T_{q s} T_{qs'}^*  \langle \Psi_{q} e^{-i H_R t} \Psi_{q}^\dagger e^{i H_R t} \rho_R \rangle_R
\end{equation}
\begin{equation} \label{equ:bcxx2-nrwa}
C^{(-)}_{ss'}(t)=\sum_{q} T_{qs} T_{qs'}^* \langle e^{-i H_R t} \Psi_{q}^\dagger  e^{i H_R t} \Psi_{q}  \rho_R \rangle_R
\end{equation}
The coupling to the DQD is assumed to be symmetric $T_{q1}=T_{q2}$. The cross coupling correlation functions are given by $C^{(\pm)}_{ss'}(t-t') = C^{(\pm)}_{ss}(t-t')$. 
The following derivations refer to Eqs.\ (\ref{equ:bcxx1},\ref{equ:bcxx2}). The same procedures have to be applied to Eqs.\ (\ref{equ:bcxx1-nrwa},\ref{equ:bcxx2-nrwa}) as well but we will not present them in detail. All the external properties of the lead are described by a single quantity, namely the spectral density $J_R(\omega)$, which can be generated by a superposition of weighted $\delta$-functions
\begin{equation} \label{equ:spectralgeneral}
J_{R}(\omega)=\sum_q \pi \vert V_q \vert^2 \delta(\omega-\omega_q).
\end{equation}
We apply a numerical decomposition of the spectral density to derive equations of motion
\begin{equation} \label{equ:spectralnum}
J_{R}(\omega)=\sum_{k=1}^m p_k  \frac{\Gamma_k^2}{(\omega -\Omega_k)^2+\Gamma_k^2}.
\end{equation}
With the complex roots of the Fermi function and of function (\ref{equ:spectralnum}), the theorem of residues applied to Eqs.\ (\ref{equ:bcxx1}, \ref{equ:bcxx2}) results in\cite{sven06_1}
\begin{eqnarray} \label{bath12dec}
C^{(+)}_{\sigma \sigma}(t)&= &\sum_{k=1}^m p_k \Gamma_k
\left(n_F(-\Omega_k^- +E_F) e^{-i\Omega_k^- t} \right) \nonumber \\
& &-\frac{2i}{\beta}  \sum_k^{m'} J_{R}(\nu_k^\ast) e^{-i \nu_k^\ast t}
\end{eqnarray}
\begin{eqnarray} \label{bath21dec}
(C^{(-)}_{\sigma \sigma})(t)&=& \sum_{k=1}^m p_k \Gamma_k
\left(n_F(\Omega_k^+-E_F)  e^{i\Omega_k^+ t} \right) \nonumber \\
& &-\frac{2i}{\beta} \sum_k^{m'} J_{R}(\nu_k) e^{i \nu_k t}
\end{eqnarray}
for $\sigma=\uparrow$ and $\sigma=\downarrow$. We use the abbreviations $\Omega_k^+=\Omega_k+i \Gamma_k$, 
$\Omega_k^-=\Omega_k-i \Gamma_k$ and the Matsubara frequencies $\nu_k$
given by $\nu_k=i\frac{2\pi k + \pi}{\beta} +E_F$. In general, one has to take an infinite number of Matsubara frequencies into account, but it was demonstrated that the summation can be truncated\cite{sven06_1,klei04a,meie99}. From Eqs.\,(\ref{bath12dec},\ref{bath21dec}), we can write the correlation functions as
\begin{equation}
C^{(\pm)}_{\sigma \sigma}(t)=\sum_{k=1}^{m+m'}  a^{(\pm)}_k e^{\gamma^{(\pm)}_k t}.
\end{equation}
The same decomposition can be derived for $C^{(\pm)}_{ss'}(t)$: 
\begin{equation}\label{specspec}
C^{(\pm)}_{ss'}(t)=\sum_{k=1}^{m+m'}  a^{(\pm)}_k e^{\gamma^{(\pm)}_k t}.
\end{equation}

\end{document}